\documentclass{PoS}
\newcommand{\be}{\begin{equation}}
\newcommand{\ee}{\end{equation}}
\newcommand{\bra}{\langle}
\newcommand{\ket}{\rangle}
\newcommand{\bea}{\begin{eqnarray}}
\newcommand{\eea}{\end{eqnarray}}
\newcommand{\dis}{\displaystyle}

\title{Simulations of one-flavor QCD at finite temperature by RHMC}

\ShortTitle{Simulations of one-flavor QCD at finite temperature by RHMC}

\author{\speaker{Tetsuya Takaishi}\\
        Hiroshima University of Economics\\
        Hiroshima 731-0192  JAPAN\\
        E-mail: \email{takaishi@hiroshima-u.ac.jp}}

\author{Atsushi Nakamura\\
        Hiroshima University\\
        Higashi-Hiroshima 739-8521 JAPAN \\
         E-mail: \email{nakamura@hiroshima-u.ac.jp}}

\abstract{We simulate one-flavor QCD with standard Wilson fermions 
at finite temperature by the rational hybrid Monte Carlo algorithm. 
In the heavy quark region when we decrease the quark mass
there is an endpoint which terminates the first order phase transition.
We try to locate it by calculating the Binder 
cumulant of the Polyakov loop norm. We estimate the end-point to be $\kappa_c \sim 0.07-0.08$.
          }

\FullConference{The XXV International Symposium on Lattice Field Theory\\
		 July 30-4 August 2007\\
		 Regensburg, Germany}

\begin{document}
\section{Introduction}
Over many years,
lattice QCD simulations have been done mainly for zero and two flavors of fermions 
because of the algorithmic difficulty of simulating odd flavors.
Recently, though, several algorithms have been developed for odd-flavor simulations.
A new comer is the rational hybrid Monte Carlo algorithm (RHMC) 
in which a fractional power of the fermion matrix needed to simulate odd fermions 
is given by a rational approximation\cite{RHMC}. 
The advantage of the RHMC is that 
the approximation error can be made small with a low approximation degree.

One-flavor QCD had not been seriously studied.
However, 
motivated by
theoretically interesting properties 
( one-flavor QCD has no pion and is expected to have no chiral symmetry )
a study of the one-flavor spectroscopy has now been started\cite{nf1HADRON}. 

One-flavor QCD at finite temperature was studied some time ago\cite{nf=1,nf=1DBW2}.
Alexandrou {\it et al.} used the multi-boson algorithm to simulate one-flavor QCD 
and mapped a rough phase diagram of the one-flavor QCD in the heavy quark region.
In order to locate the end-point accurately, they also used the effective 3D Potts model 
and estimated $\kappa_c \sim 0.08$\cite{nf=1}.

In this study we use the RHMC to simulate one-flavor QCD
and try to locate the end-point by the Binder cumulant of the Polyakov loop norm.

\section{One-flavor algorithms}

The lattice QCD partition function is given by
\be
Z= \int [dU] \det D(U)^{n_f} e^{-S_g(U)},
\ee
where $D(U)$ is the fermion matrix, $n_f$ the number of flavors and $S_g(U)$ the gauge action.
We use the standard Wilson fermion and the standard Wilson gauge action. 
With this partition function the expectation value of an operator $\Omega$ is given by
\be
\bra \Omega \ket
= \frac1{Z} \int [dU] \Omega[U]  \det D(U)^{n_f} e^{-S_g(U)}.
\ee
For multiples of even-flavor the fermionic determinant ( here for 2 flavors as an example ) 
is expressed as
\be
\det D^\dagger D =\int D\phi^\dagger D\phi  e^{-\phi^\dagger ( D^\dagger D)^{-1} \phi} 
= \int  D\phi^\dagger D\phi   e^{-\eta^\dagger  \eta},
\ee
where $\eta={D^\dagger }^{-1} \phi$.
The key ingredient of the conventional hybrid Monte Carlo (HMC)\cite{HMC} is 
that the fermionic action, $S_f=\phi^\dagger ( D^\dagger D)^{-1} \phi$
is 
manifestly positive. 
This positive form of the action can not be easily realized for odd-flavors.
For this positive fermionic action one can easily update $\phi$ using the heat-bath method, i.e. 
$\phi={D^\dagger } \eta$, where $\eta$ is drawn by $P(\eta)\sim e^{-\eta^\dagger  \eta}$.

If we use the identity of $\det D^{n_f} = n_f tr\log D$ we can make simulations of 
any numbers of flavors using the R-algorithm\cite{R-algo} which is,
however, an inexact algorithm introducing errors to the results.
When we calculate the fermionic forces in the R-algorithm,
the calculations of $tr(D^{-1}D^\prime)$ appear. 
Such calculations are numerically very costly and usually they are estimated approximately using 
the random noise method, which introduces the approximation errors.
These errors can not be removed completely.
The order of the errors is $O(\delta t^2)$, where $\delta t$ is the step-size. 
In order to obtain the exact results from the R-algorithm,
one needs to extrapolate them to the zero step-size limit, i.e. $\delta t \rightarrow 0$.    
 
The exact algorithms of odd flavors can be constructed using L\"{u}scher's idea\cite{Luescher}.
He approximates the inverse of the fermion matrix using the following  polynomial.
\be
\frac1{D} \approx \Pi_{i=1}^{n} (D-z_i),
\label{POL}
\ee
where $z_i$ are the roots of the polynomial.
With this polynomial a multi-boson algorithm which is algorithmically very different from the HMC can be 
constructed.
Originally it was applied for even flavors
and later generalized to any numbers of flavors\cite{MULTI1,MULTI2}. 
Using the $\gamma_5$ hermiticity of $D$, $D=\gamma_5 D^\dagger \gamma_5$, 
the determinant of a single $D$ for $n_f=1$ is 
written as
\bea
\det D &  \approx & \det[T^\dagger_{n/2}(D) T_{n/2}(D)]^{-1},   \\
      &  \propto  & \int \Pi_{i=1}^{n/2} D\phi_i^\dagger D\phi_i e^{-\sum_{i=1}^{n/2}\phi^\dagger_i(D-z_i)^\dagger (D-z_i) \phi_i}, \\
      &   = &      \int \Pi_{i=1}^{n/2} D\phi_i^\dagger  D\phi_i e^{-\sum_{i=1}^{n/2}\eta_i^\dagger \eta_i},
\eea
where $T_{n/2}(D)=\Pi_{i=1}^{n/2}(D-z_i)$ and $\eta_i = (D-z_i) \phi_i$.
Each bosonic action $\dis\phi^\dagger_i(D-z_i)^\dagger (D-z_i) \phi_i$ is now positive
and $\phi_i$ are updated using the heat-bath method.

The same idea of using the polynomial can also be used in the framework of the HMC algorithm.
de~Forcrand and Takaishi first used L\"{u}scher's polynomial for the HMC 
in order to reduce the computational cost of the algorithm\cite{FAST_FERMION}.
With the help of the polynomial, an odd-flavor HMC algorithm is constructed\cite{ODD_HMC} as follows.
Using the $\gamma_5$ hermiticity 
of
$D$, the determinant of $D$ is written as
\bea
\det D &  \approx & \det [T^\dagger_{n/2}(D) T_{n/2}(D)]^{-1}, \\
      &  \propto  & \int D\phi^\dagger D\phi e^{-\phi^\dagger [T^\dagger_{n/2}(D) T_{n/2}(D)]\phi}, \\
      &   = &      \int D\phi^\dagger D\phi e^{-\eta^\dagger \eta},
\eea
where $T_{n/2}(D)=\Pi_{i=1}^{n/2}(D-z_i)$ and $\eta =  T_{n/2}(D)\phi$.
Thus, as in the conventional HMC algorithm, the fermionic action $S_f=\phi^\dagger [T^\dagger_{n/2}(D) T_{n/2}(D)]\phi$
is made to be positive and $\phi$ is updated using the heat-bath method.

L\"{u}scher's polynomial is not nessesarily the optimal one.
Instead, one can find a rational approximation with more accuracy.
Clark and Kennedy constructed the HMC algorithm with a rational approximation\cite{RHMC},
which will be described briefly in the next section.

\section{Rational Hybrid Monte Carlo}
 
In the RHMC the fermion determinant is 
rewritten as
\be 
\det M^\alpha \propto \int D\phi^\dagger D\phi e^{-\phi^\dagger M^{-\alpha} \phi} =\int  D\phi^\dagger D\phi e^{-\phi^\dagger r(M)_n^2 \phi},
\ee
where $\alpha=n_f/2$, $M=(D^\dagger D)$ and the rational approximation $r_n(M)^{-1}$ is given by
\be
r_n(M)^{-1}=\alpha_0 + \sum_{i=1}^n \frac{\alpha_i}{(M-\beta_i)}.
\ee
In this way, the fermionic action $S_f=\phi^\dagger r_n(M)^2 \phi$ is made to be positive.
Thus the framework of the HMC algorithm can be used with this approximation.

The operations such as $r(M)_n^{-1}\eta$ including $(M-\beta_i)^{-1}\eta$ can be done by
the multi-shift solver\cite{Multi-shift}.
The computational cost of this operation is expected to be similar to the HMC\cite{RHMC}.

\begin{figure}
\center{
\includegraphics[width=.6\textwidth,clip]{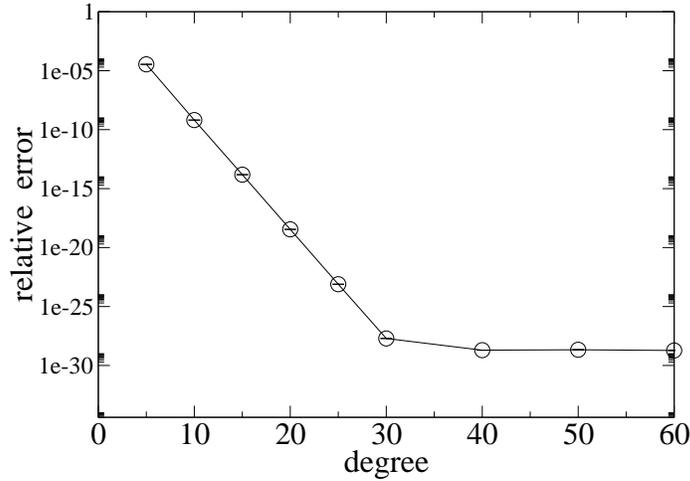}
\caption{The relative error as a function of the approximation degree.}
\label{fig1}
}
\end{figure}

The coefficients of the rational approximation are numerically calculated by the Remez method\cite{Remez}.
The rational approximation error can be made small with decreasing order $n$.
Typically, $O(20)$ or less gives the maximum relative error that is good to machine precision\cite{RHMC}.
Fig.\ref{fig1} shows an example of the relative errors calculated on $20^3 \times 4$ lattices.
We calculate $\displaystyle |r_n(M)^{-1}\phi-r_{n_{max}}(M)^{-1}\phi|^2/|r_{n_{max}}(M)^{-1}\phi|^2$ 
as a function of $n$. Here $\phi$ is random gaussian noise vectors 
and $n_{max}$ is set to 60.
The results shown on Fig.1 are averages taken over 10 configurations
at $\beta=5.63$ and $\kappa=0.12$. 
The error bars are smaller than the symbols.
We see that with increasing $n$ the relative error decreases quickly. 
The relative error with $n=20$ is already very small, less than $10^{-15}$.
For our simulations we take $n=20$ or 25( for the largest lattice $20^3\times 4$ ).

Here, note that the low approximation degree itself does not
mean that the cost of the RHMC is small.
The rational approximation in the RHMC contains 
solver calculations such as $D^{-1}\phi$.
Roughly speaking the cost of the RHMC is proportional to  
that of the solver calculations.
On the other hand, in the polynomial HMC the calculation of $D^{-1}\phi$
is replaced with the calculation by the polynomial approximation.
Therefore, the cost of the polynomial HMC is directly 
proportional to the number of the approximation degree.

\begin{figure}
\hspace{2mm}
\center{
\includegraphics[width=.6\textwidth,clip]{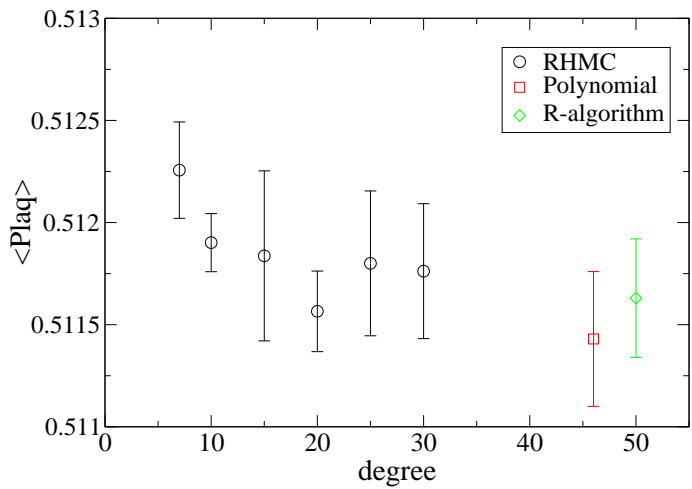}
\caption{Plaquette values calculated on a $6^4$ lattice at $\beta=5.45$ and $\kappa=0.16$
as a function of the approximation degree. The results from the polynomial HMC and the R-algorithm 
are taken from \cite{ODD_HMC}.}
\label{fig2}
}
\end{figure}

Fig.\ref{fig2} shows the plaquette values obtained from $n_f=1$ simulations 
on a $6^4$ lattice at $\beta=5.45$ and $\kappa=0.160$  as a function of the approximation degree.
The results from the RHMC are consistent with those from the R-algorithm and the polynomial HMC
unless the approximation degree is very small, like $n\sim 5$.

\section{One-flavor simulations}
\begin{figure}
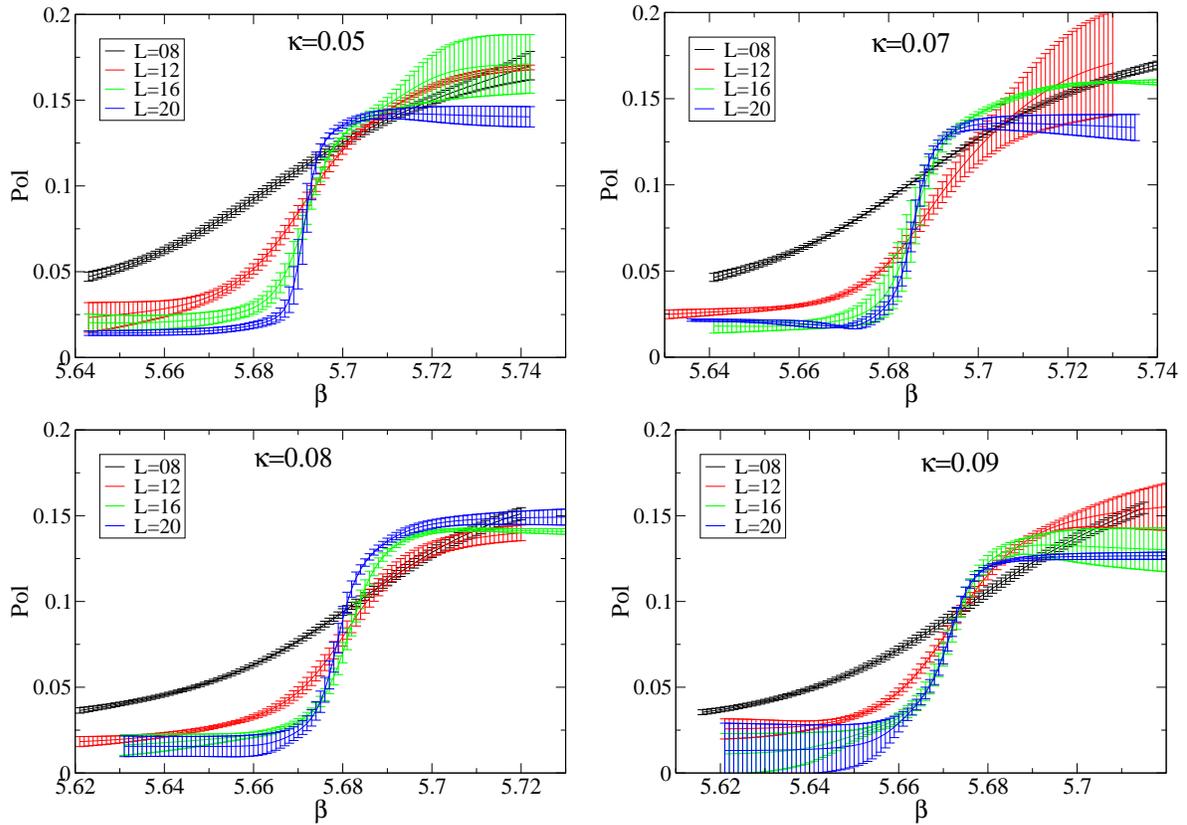

\center{
\begin{tabular}{cc}
\includegraphics[width=.49\textwidth,clip]{re_pol_beta_k050.eps} &
\includegraphics[width=.51\textwidth,clip]{re_pol_beta_k070.eps} \\
\includegraphics[width=.49\textwidth,clip]{re_pol_beta_k080.eps} &
\includegraphics[width=.49\textwidth,clip]{re_pol_beta_k090.eps} \\
\end{tabular}
\caption{Polyakov loop norm for different lattice sizes as a function of $\beta$.}
\label{fig3}
}
\end{figure}

We simulated one-flavor QCD on  $L^3 \times 4$ lattices with $L=8,12,16$ and 20.
We used an improved integrator, 2MN integrator, for molecular dynamics simulations 
in the RHMC. The 2MN integrator is $50\%$ faster than the conventional 2nd order
leap frog integrator\cite{2MN}. 
The step-sizes are set to the values that give acceptances from $60\%$ to $70\%$.
These acceptances are shown to be optimal for any 2nd order integrator\cite{HOHMC}.
We generated about $2\times 10^5$ ($1\times 10^5$) trajectories for L=8 ($12\sim 20$).

Fig.\ref{fig3} shows the Polyakov loop norm for each $\kappa$ as a function of $\beta$.
At large $\kappa$ we see no evidence that on larger lattices 
the discontinuity across $\beta_c$ will be pronounced.
On the other hand, at small $\kappa$ the discontinuity 
appears for larger lattices,  
which may show evidence of first order phase transition.
\begin{figure}
\center{
\includegraphics[width=.6\textwidth,clip]{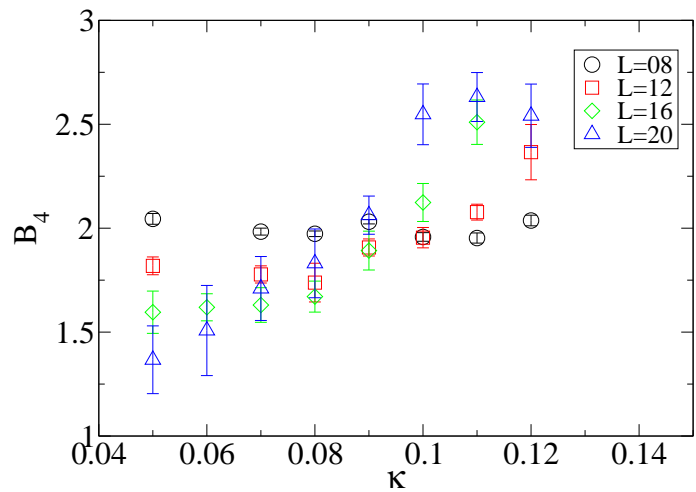}
\caption{Binder cumulant of the Polyakov loop norm for various lattice sizes
as a function of $\kappa$.}
\label{fig4}
}
\end{figure}

We calculate the Binder cumulant $B_4$ defined by
\be
B_4= \frac{\bra \Omega^4\ket}{\bra \Omega^2\ket^2}.
\ee
$B_4$ takes 3 or 1 for the first order phase transition or crossover.
At the critical point $B_4$ takes a certain value dependent on the universality class.
For 3d Ising universality $B_4$ is 1.604\cite{Ising}.
Fig.\ref{fig4} shows $B_4$ for the Polyakov loop norm as a function of $\kappa$.
The results are still noisy and the clear critical point where all curves intersect 
is not determined yet. However we estimate it to be $\kappa_c = 0.07\sim0.08$.  

\section{Summary}
The one-flavor QCD simulations are done by the RHMC.
The relative error of the rational approximation 
can be made small using a low approximation degree such as $n\sim 20$. 
We compare the plaquette results from the RHMC with 
those from the polynomial HMC and the R-algorithm.
The results are consistent with those from the polynomial HMC and the R-algorithm.

We calculate the Binder cumulant $B_4$ of the Polyakov loop norm
on $L^3 \times 4$ lattices with $L=8,12,16$ and 20.
The $B_4$ curves are expected to intersect at the critical point $\kappa_c$
and we estimate roughly $\kappa_c \sim 0.07-0.08$. 

In future in order to locate the end-point accurately we plan to use a larger lattice
and to improve our data statistics.

\acknowledgments
The numerical calculations were carried out on an SX8 at the YITP in Kyoto University,
an SX8 at the RCNP in Osaka University and an SX6 at the Institute of Statistical Mathematics.
The work was supported in part by Hiroshima University of Economics,
and
by Grants-in-Aid for Scientific Research
from the Monbu-Kagaku-sho (No. 17340080).

\end{document}